\documentclass[conference]{IEEEtran}
\usepackage{graphicx}
\usepackage{amssymb}
\usepackage{amsmath}
\usepackage{cite}
\usepackage{algorithm}
\usepackage{color}
\usepackage{epstopdf}
\usepackage[applemac]{inputenx}
\usepackage{algpseudocode}

\newcommand{\B}[1]{{\pmb{#1}}}

\ifCLASSINFOpdf 
\else
\fi
\begin{document}

\title{Distributed Algorithms for Peak Ramp Minimization Problem in Smart Grid}

\author{\IEEEauthorblockN{Hung Khanh Nguyen\IEEEauthorrefmark{1}, Amin Khodaei\IEEEauthorrefmark{2} and Zhu Han\IEEEauthorrefmark{1}}
	\IEEEauthorblockA{\IEEEauthorrefmark{1}Department of Electrical and Computer Engineering,
		University of Houston, Houston, Texas 77004}
	\IEEEauthorblockA{\IEEEauthorrefmark{2}Department of Electrical and Computer Engineering,
		University of Denver, Denver, CO 80210}
}



\maketitle

\begin{abstract}
The arrival of small-scale distributed energy generation in the future smart grid has led to the emergence of so-called prosumers, who can both consume as well as produce energy. By using local generation from renewable energy resources, the stress on power generation and supply system can be significantly reduced during high demand periods. However, this also creates a significant challenge for conventional power plants that suddenly need to ramp up quickly when the renewable energy drops off. In this paper, we propose an energy consumption scheduling problem for prosumers to minimize the peak ramp of the system. The optimal schedule of prosumers can be obtained by solving the centralized optimization problem. However, due to the privacy concerns and the distributed topology of the power system, the centralized design is difficult to implement in practice. Therefore, we propose the distributed algorithms to efficiently solve the centralized problem using the alternating direction method of multiplier (ADMM), in which each prosumer independently schedules its energy consumption profile. The simulation results demonstrate the convergence performance of the proposed algorithms as well as the capability of our model in reducing the peak ramp of the system.
\end{abstract}

\IEEEpeerreviewmaketitle

\section{Introduction}
%

%
%
%
%
%
%
%

%
%
%
%
%
%
%

The smart grid concept has been proposed as the essential element to facilitate the interaction between different domains and entities involved in the current power network \cite{Zhuhan:Smartgridbook}. By utilizing modern information and communication technologies, smart grid is expected to be one of the key enablers to improve reliability, resiliency, flexibility, and efficiency of the electric delivery system \cite{Xifang_smargridsurvey}. In addition, the recent deployment of small-scale distributed energy generation such as customer-sited photovoltaic systems has led to the emergence of so-called prosumers. Prosumers are entities that can both consume as well as produce energy, mainly from renewable energy resource. By using local renewable energy resource during high demand periods, distributed generation has been advocated as a promising solution to improve security of supply as well as reduce environmental impacts \cite{hill_battery}. Many energy and environmental policy initiatives have been realized to increase distributed generation and ensure the efficient and sustainable use of natural resource. According to the Renewable Portfolio Standard \cite{denholm_overgeneration}, 33\% of California's electricity is required to come from renewable resources by 2020, which is expected to reduce greenhouse gas emission to 1990 levels. 

%
%

\begin{figure}[t]
	\centering
	\includegraphics[width=0.4\textwidth]{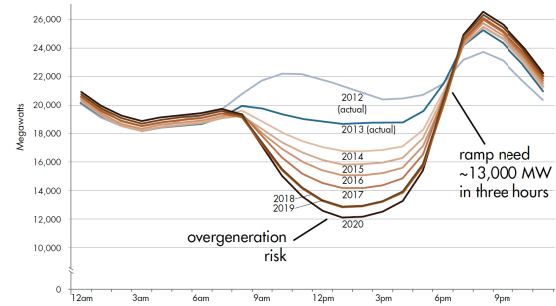}
	\caption{
		Peak ramp problem of the system net load in California \cite{caiso_duck}.
	}
	\label{fig:duckissue}\vspace{-5mm}
\end{figure}

Although deploying of distributed generation such as massive amounts of customer-sited photovaltaic systems has brought environmental and economical benefits, it also creates challenges for grid operations. Due to the intermittent nature of renewable generation, the grid operator must be able to either drive down the generation output when renewable generation units start producing power, or ramp up generation when renewable generation shuts down. For example, Fig. \ref{fig:duckissue} shows the effect of distributed generation on the net load of the system in California \cite{caiso_duck}, which is calculated by taking the forecasted load and subtracting the forecasted electricity production from variable generation resources, mainly from solar generation. Due to the large amount of solar power available during midday when distributed photovoltaic generation is at the highest capacity, the net load of the system is pulled down to extremely low levels. Then, later in the day when solar generation is declining, the net load of the system ramps up dramatically. It is projected that the system operators in California must be able to ramp up $13000$ MW in three hours to satisfy customers demand. Tremendous research and industry efforts have investigated demand side management programs, which control the energy consumption at the customer side to make power grids more reliable and robust. \cite{Amir-Hamed,hung_distrbuted,song_demandside,christianIbars} propose incentive mechanisms to induce customers to reschedule their demand in response to power supply conditions to reduce energy payment. The works in \cite{hung_peak,caron,chen_aninnovative} propose real-time pricing to reduce the peak-to-average load ratio of the system.

In this paper, we study an energy consumption scheduling problem for prosumers in the future smart grid by taking into account the impact of distributed solar generation on the net load of the system. The objective of the optimization problem is to minimize the peak ramp of the system. The formulated problem can be solved efficiently by a central controller with complete information from prosumers. However, due to the distributed topology of the power system, the centralized method is difficult to implement in practice. Therefore, we propose two distributed algorithms to obtain the global optimal solution using the alternating direction method of multiplier (ADMM) \cite{boyd_admm}, in which each prosumer independently schedules its energy consumption profile.

The remainder of this paper is organized as follows. We explain the model for energy consumption of prosumers in Section \ref{sec:SystemModel}, and formulate the peak ramp minimization problem for the system in Section \ref{sec:problem}. Section \ref{sec:algo} provides distributed algorithms to solve the energy consumption scheduling problem. Simulation results are presented in Section \ref{sec:simulation}, and Section \ref{sec:conclusions} concludes the paper.

\section{System Model} \label{sec:SystemModel}
%

%
%
\begin{figure}[t]
	\centering
	\includegraphics[width=0.45\textwidth]{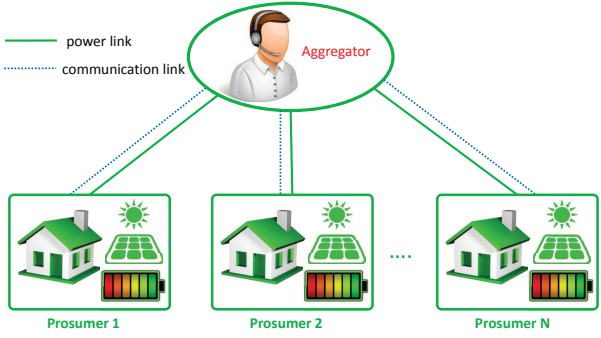}\vspace{-1mm}
	\caption{
		The model of power system with prosumers and the aggregator.
	}
	\label{fig:system}\vspace{-5mm}
\end{figure}

We consider a distribution grid consisting of $N$ prosumers, denoted by the set $\mathcal{N} \triangleq \{1, 2, \ldots, N\}$, each of which is equipped with a renewable generation unit and an energy storage system, as illustrated in Fig. \ref{fig:system}. Note that our model can be applied to systems that have only a subset of prosumers have energy storage systems or distributed generation units by setting the storage capacity to zero for any prosumer out of the subset. Each prosumer acquires energy from the grid operator and local resource (i.e., from renewable energy and storage system) to supply its demand, which consists of elastic portion and inelastic portion. The elastic portion is the fixed amount of energy consumption during a fixed period of time, while the elastic portion has the shifting flexibility to adjust the energy consumption across time.

The energy scheduling problem is considered in a one-day period which is divided into a set of $T$ equal time slots, denoted by $\mathcal{T} \triangleq \{1, 2, \ldots, T\}$ . For each prosumer $n \in \mathcal{N}$, let $\B{P}_n \triangleq \{ P_n[1], P_n[2], \ldots, P_n[T] \}$ be the fixed energy consumption vector for inelastic demand over the scheduling horizon. We also define $e_n[t]$ as the energy consumption of elastic demand at time slot $t$, which must satisfy the minimum and maximum energy consumption levels
\begin{align}
	e_n^{min} \leq e_n[t] \leq e_n^{max}.
\end{align}
The amount of energy consumption for elastic demand must be fulfilled at the end of the scheduling horizon, which can be expressed as the energy load balance constraint as follow
\begin{align}
\sum_{t=1}^{T} e_n[t] = E_n,
\end{align}
where  $E_n$ is the predetermined total daily energy demand of elastic load.


At each time slot, a prosumer can either discharge energy from its storage system to supply local demand or charge energy for the storage system. Let $x_n[t]$ and $y_n[t]$ be the amount of energy charging and discharging for the energy storage system of prosumer $n$ at time slot $t$, respectively, which satisfy the maximum charging $x_n^{max}$, and discharging $y_n^{max}$ as
\begin{align}
	 0 \le x_n[t] \le x_n^{max},  \\
	 0 \le y_n[t] \le y_n^{max}. 
\end{align}
Then, the energy level of the storage system at time slot $t$ can be determined as
\begin{align}
s_n[t+1] = s_n[t] + \beta_n^c x_n[t] - y_n[t],
\end{align}
where $\beta_n^c$ is the charging efficiency. The energy level in the storage system must be greater than zero and less than the storage capacity as the following constraint
\begin{align}
	0 \leq s_n[t] \leq B_n^{cap},
\end{align}
where $B_n^{cap}$ is the maximum storage capacity.

Let $W_n[t]$ be the amount of renewable energy (i.e., wind or solar energy) that prosumer $n$ obtains at time $t$. We assume that prosumers can predict renewable generation for the scheduling horizon using historical data or machine learning methods \cite{sharma_predicting}. Then we can calculate the remaining energy that prosumer $n$ acquires from the main grid to fully satisfy its demand as
\begin{align}
	d_n[t] = P_n[t] + e_n[t] + x_n[t] - \beta_n^d y_n[t] - W_n[t],
\end{align}
where $\beta_n^d$ is the discharging efficiency. Based on the energy request from all prosumers in each time slot, the total net load that the system needs to meet for all prosumers can be calculated as
\begin{align}
	l[t] = \sum_{n=1}^{N} d_n[t].
\end{align}
We further define net load profile vector of the system as
\begin{align}
	\B{l} \triangleq \{l[1], l[2], \ldots, l[T] \}.
\end{align}
Then the ramp between two consecutive time slots, $t$ and $t-1$, can be calculated as
\begin{align}
	r[t] = l[t] - l[t-1] = \sum_{n = 1}^N d_n[t] - \sum_{n=1}^N d_n[t-1],
\end{align}
where the ramp for the case $r[1]$ can be calculated as the difference between the net load at time slot $t=1$ and the net load at the last time slot of previous day, which is assumed to be known at the beginning of the optimization process.

Then the ramping vector of the system over the scheduling horizon is
\begin{align}
\B{r} \triangleq \{r[1], r[2], \ldots, r[T] \}.
\end{align}

\section{Problem Formulation} \label{sec:problem}
In this section, we formulate the peak ramp minimization problem for the system. Based on the ramping vector defined in the previous section, we can find the maximum ramp of the system over the scheduling horizon as the infinity-norm of the ramping vector
\begin{align}
\|\B{r}\|_{\infty} = \max \{ |r[t]|: t = 1,2, \ldots,T \}.
\end{align}
Then, each prosumer schedules its energy consumption and storage profile to minimize the peak ramp of the system as the following optimization problem 
\begin{align}
&\min && \|\B{r}\|_{\infty} \label{opt:min_norm}\\
& \text{s.t.}
&& r[t] = \sum_{n=1}^N d_n[t] - \sum_{n=1}^{N} d_n[t-1], \forall t,   \label{const:start_minnorm}\\
&&& d_n[t] = P_n[t] + e_n[t] + x_n[t] \nonumber \\ &&& \qquad \qquad  - \beta_n^d y_n[t] - W_n[t], \forall t, \forall n,  \label{const:prosumer_netload} \\
&&& \sum_{t=1}^{T} e_n[t]  = E_n, \forall n,  \\
&&& s_n[t+1] = s_n[t] + \beta_n^c x_n[t] - y_n[t], \forall t, \forall n,  \\
&&& 0 \leq s_n[t] \leq B_n^{max}, \forall t, \forall n,  \\
&&& 	e_n^{min} \leq e_n[t] \leq e_n^{max}, \forall t, \forall n, \\
&&& 0 \le x_n[t] \le x_n^{max}, \forall t, \forall n, \\
&&& 0 \le y_n[t] \le y_n^{max}, \forall t, \forall n, \label{const:end_minnorm}\\
& \text{variables:}
&& \{\B{e}_n, \B{x}_n, \B{y}_n, \B{d}_n \}_{\forall n}, \B{r}. \nonumber
\end{align}

The problem in \eqref{opt:min_norm}-\eqref{const:end_minnorm} is difficult to solve in its original form. We transform it into an equivalent optimization problem by introducing an auxiliary variable

\begin{align}
&\min && \Gamma \label{opt:min_gamma} \\
& \text{s.t.}
&&  - \Gamma \le r[t] \le \Gamma, \forall t \in \mathcal{T},  \nonumber \\
&&& \text{Constraints} \, \eqref{const:start_minnorm}-\eqref{const:end_minnorm}, \nonumber \\
& \text{variables:}
&& \{\B{e}_n, \B{x}_n, \B{y}_n, \B{d}_n \}_{\forall n}, \B{r}, \Gamma. \nonumber
\end{align}

The problem in \eqref{opt:min_gamma} can be solved using the convex optimization technique \cite{boyd_convex}. However, we need to have a central controller to collect all information of prosumers, which is difficult to implement in practice due to the distributed topology of the power network as well as privacy concerns of prosumers. In the next section, we propose the distributed algorithms to solve the optimization problem in \eqref{opt:min_gamma} in which each prosumer individually solves its own optimization problem to achieve the global optimal solution for the system.
\section{Distributed Algorithms} \label{sec:algo}
In this section, we propose two distributed algorithms to achieve the global optimal solution for the optimization problem in \eqref{opt:min_gamma}. 
\subsection{Synchronous Distributed Algorithm}
In this subsection, we use the ADMM decomposition method to propose a synchronous distributed algorithm for the problem in \eqref{opt:min_gamma}. This algorithm decompose the original problem into $N + 1$ subproblems, which can be solved by the aggregator and prosumers distributively.

The optimization problem in \eqref{opt:min_gamma} has a large number of constraints. However, we realize that constraints \eqref{const:prosumer_netload}-\eqref{const:end_minnorm} are separated for each prosumer. The only constraints in \eqref{const:start_minnorm} are coupled over different prosumers. In order to make constrains in \eqref{const:start_minnorm} to be separable for each prosumer, we define auxiliary variables
\begin{align}
	d_n[t] = \hat{d}_n[t], \forall t \in \mathcal{T}, \forall n \in \mathcal{N},
\end{align}
where each auxiliary variable $\hat{d}_n[t]$ can be interpreted as the local copy of $d_n[t]$. Then the constraint in \eqref{const:start_minnorm} can be rewritten as
\begin{align}
	r[t] = \sum_{n=1}^N \hat{d}_n[t] - \sum_{n=1}^{N} \hat{d}_n[t-1], \forall t.
\end{align}
Moreover, to facilitate for presentation, we define the feasible set for each prosumer as
\begin{align}
\mathcal{F}_n = \{ \{ \B{d}_n, \B{e}_n, \B{x}_n, \B{y}_n \} | \eqref{const:prosumer_netload}-\eqref{const:end_minnorm} \}. \nonumber
\end{align}
Then the problem in \eqref{opt:min_gamma} can be rewritten as 
\begin{align}
&\min && \Gamma  \label{opt:min_gamma_sep}\\
& \text{s.t.}
&&  - \Gamma \le r[t] \le \Gamma, \forall t \in \mathcal{T},  \nonumber \\
&&& r[t] = \sum_{n=1}^N \hat{d}_n[t] - \sum_{n=1}^{N} \hat{d}_n[t-1], \forall t, \nonumber \\
&&& d_n[t] = \hat{d}_n[t], \forall t \in \mathcal{T}, \forall n \in \mathcal{N}, \nonumber \\
&&& \{ \B{d}_n, \B{e}_n, \B{x}_n, \B{y}_n \} \in \mathcal{F}_n, \forall n, \nonumber \\
& \text{variables:}
&& \{\B{e}_n, \B{x}_n, \B{y}_n, \B{d}_n, \B{\hat{d}}_n \}_{\forall n}, \B{r}, \Gamma. \nonumber
\end{align}

The augmented Lagrangian function of the problem in \eqref{opt:min_gamma_sep} is given by \cite{boyd_admm}
\begin{align}
\mathcal{L} & = \Gamma + \sum_{n=1}^{N} \sum_{t=1}^T \mu_{n,t} (\hat{d}_n[t] - d_n[t]) + \dfrac{\rho}{2} \sum_{n=1}^N \| \B{\hat{d}}_n- \B{d}_n \|^2 \nonumber \\
& = \Gamma + \sum_{n=1}^{N} \sum_{t=1}^T \mu_{n,t} \hat{d}_n[t]  -  \sum_{n=1}^{N} \sum_{t=1}^T \mu_{n,t} d_n[t] \nonumber \\
  & \qquad \qquad \qquad \qquad \qquad  + \dfrac{\rho}{2} \sum_{n=1}^N \| \B{\hat{d}}_n- \B{d}_n \|^2 , \label{eq:langrangian} 
\end{align}
where $\{\mu_{n,t}\}_{\forall n, \forall t}$ is the Lagrangian multiplier, and $\rho > 0$ is a penalty parameter.

Defining the primal variable $\B{u} = \{ \{\B{\hat{d}}_n\}_{\forall n}, \B{r}, \Gamma  \}$, which is the decision variable for the system aggregator, and $\B{u} = \{ \{\B{d}_n, \B{x}_n, \B{y}_n, \B{e}_n  \}_{\forall n}\}$ is the decision variable for prosumers. Then we can use ADMM decomposition technique to solve the optimization problem in \eqref{opt:min_gamma_sep} in an iterative procedure. Particularly, at the $k$-th iteration, the primal variables and dual variables can be sequentially updated as
\begin{align}
\B{u}^{[k+1]} & = \arg \min 	\mathcal{L}(\B{u}, \B{v}^{[k]}, \B{\mu}^{[k]} ), \label{opt:primal_aggregator} \\
\B{v}^{[k+1]} & = \arg \min 	\mathcal{L}(\B{u}^{[k+1]}, \B{v},\B{\mu}^{[k]} ),  \label{opt:primal_prosumer} \\
\B{\mu}^{[k+1]} & = \B{\mu}^{[k]} + \rho \left ( \B{\hat{d}}^{[k+1]} - \B{d}^{[k+1]} \right).  \label{opt:dual}
\end{align}

Based on the Lagrangian function in \eqref{eq:langrangian}, we decompose the problem in \eqref{opt:min_gamma_sep} into $N + 1$ optimization problems. The first subproblem is associated with the primal variables for the aggregator as in \eqref{opt:primal_aggregator} 
\begin{align}
&\min && \Gamma + \sum_{n=1}^{N} \sum_{t=1}^T \mu_{n,t} \hat{d}_n[t] +  \dfrac{\rho}{2} \sum_{n=1}^N \| \B{\hat{d}}_n- \B{d}_n \|^2 \label{opt:local_aggregator} \\
& \text{s.t.}
&& - \Gamma \leq r[t] \leq \Gamma, \forall t,  \nonumber\\
&&& r[t] = \sum_{n=1}^{N} \hat{d}_n[t] - \sum_{n=1}^{N} \hat{d}_n[t-1], \forall t, \nonumber\\
& \text{variables:}
&& \Gamma, \B{r}, \{\B{\hat{d}}_n\}_{\forall n}. \nonumber
\end{align}

The remaining $N$ subproblems are associated with variables for each prosumer and corresponding to primal variables update in \eqref{opt:primal_prosumer}. Each prosumer $n \in \mathcal{N}$ solves its own problem as
\begin{align}
&\min && -  \sum_{t=1}^T \mu_{n,t} d_n[t] + \dfrac{\rho}{2} \| \B{\hat{d}}_n- \B{d}_n \|^2 \label{opt:local_prosumer} \\
& \text{s.t.}
&& \{ \B{d}_n, \B{e}_n, \B{x}_n, \B{y}_n \} \in \mathcal{F}_n. \nonumber 
\end{align}

\begin{algorithm}[h]
	\caption{Synchronous ADMM}\label{alg:parallel}
	\begin{algorithmic}[1]			
		\State initialize: $k = 0$, $\B{\mu}_n = 0, \forall n$ 		
		\Repeat 
		\State \textbf{At the aggregator:}
		\Repeat
		\State wait
		\Until{receive updates $\B{\mu}_n, \B{d}_n$ from all prosumers}
		\State \quad 1) solve local problem in \eqref{opt:local_aggregator} for the optimal solution $\Gamma, \{\B{\hat{d}}_{n}\}_{\forall n}, \B{r}$
		\State \quad 2) send $\B{\hat{d}}_{n}$ to corresponding prosumer
		\State -----------------------------------------
		\State \textbf{At each prosumer:}
		\Repeat
		\State wait
		\Until{receive the update $\B{\hat{d}}_{n}$ from the aggregator}
		\State \quad 1) solve local problem in \eqref{opt:local_prosumer} for optimal solution $\{ \B{d}_n, \B{e}_n, \B{x}_n, \B{y}_n \}$ 
		\State \quad 2) update dual variables:
		\begin{align}
		\B{\mu}_{n}^{[k+1]} & = \B{\mu}_{n}^{[k]} + \rho \left ( \B{\hat{d}}_n^{[k+1]} - \B{d}_n^{[k+1]} \right),  \nonumber
		\end{align}
		\State \quad 3) send $\B{\mu}_n, \B{d}_n$ to the aggregator
		\State ------------------------------------------
		\State $k \leftarrow k+ 1$
		\Until{a stopping criterion is met}
	\end{algorithmic}
\end{algorithm}

\begin{figure}[t]
	\centering
	\includegraphics[width=0.5\textwidth]{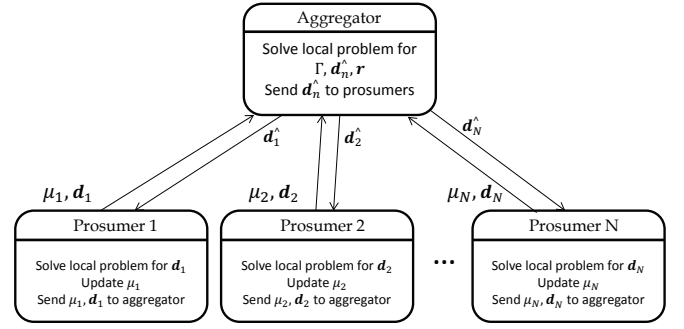}
	\caption{
		The information exchange between the aggregator and prosumers.
	}
	\label{fig:sync_info}
\end{figure}

\begin{figure}[t]
	\centering
	\includegraphics[width=0.5\textwidth]{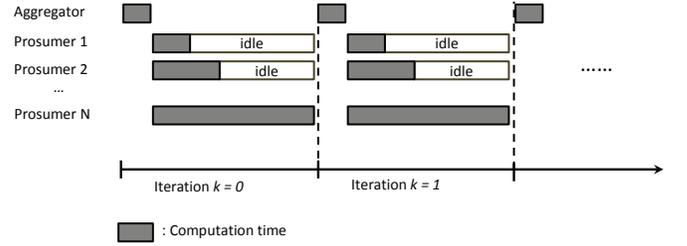}
	\caption{
		Illustration of synchronous parallel computing for Algorithm \ref{alg:parallel}. New iteration starts after the slowest prosumer finishes.
	}
	\label{fig:sync_time}
\end{figure}

The whole procedure for solving the problem in \eqref{opt:min_gamma_sep} can be summarized in Algorithm \ref{alg:parallel}. The information exchange between the aggregator and prosumers is illustrated in Fig. \ref{fig:sync_info}. However, Algorithm \ref{alg:parallel} must be performed in the synchronous fashion. Specifically, in each iteration, the aggregator has to wait until receiving all updated values from prosumers as depicted in Fig. \ref{fig:sync_time}. In this computing framework, the aggregator must wait for the slowest prosumer to finish computation before a new iteration can be triggered.

\subsection{Asynchronous Distributed Algorithm}
In this subsection, we propose a distributed algorithm that can solve the optimization problem in an asynchronous fashion by using the asynchronous parallel coordinate updates method \cite{peng_arock}. Particularly, the aggregator and prosumers do not need to wait for the slowest agent to finish computation to start a new iteration.


Consider an optimization problem with the general form as
\begin{align}
& \min && f(x) + g(y) \label{opt:generalADMM}\\
& \text{s.t.} 
&& Ax + By = b. \nonumber
\end{align}

The asynchronous parallel ADMM framework to solve the problem in \eqref{opt:generalADMM} can be performed as \cite{peng_arock}
\begin{align}
y^{k+1} & := \underset{y}{\arg \min} \,	g(y) - \langle z^k, By -b\rangle + \frac{\gamma}{2} \| By -b \|^2, \label{eq:asyn_step1}\\
w_g^{k+1} &= z^k - \gamma (By^{k+1} - b), \label{eq:asyn_step2} \\
x^{k+1} & := \underset{x}{\arg \min} \,	f(x) - \langle 2w_g^{k+1} - z^k, Ax\rangle + \frac{\gamma}{2} \| Ax \|^2, \label{eq:asyn_step3} \\
w_f^{k+1} &= 2 w_g^{k+1} - z^k - \gamma A x^{k+1}, \label{eq:asyn_step4} \\
z_n^{k+1} & = z_n^k + \eta ( w_{f,n}^{k+1}  - w_{g,n}^{k+1} ) \label{eq:asyn_step5},
\end{align}
where \eqref{eq:asyn_step2}-\eqref{eq:asyn_step5} normally can be decomposed into each agent, and \eqref{eq:asyn_step1} can be  performed in an asynchronous fashion.

By applying the procedure in \eqref{eq:asyn_step1}-\eqref{eq:asyn_step5} into \eqref{opt:min_gamma_sep}, we can solve the problem in \eqref{opt:min_gamma_sep} in asynchronous distributed fashion. Specifically, the aggregator solves the following local optimization problem
\begin{align}
&\min && \Gamma + \sum_{n=1}^N \langle \B{z}_n ,\B{\hat{d}}_n \rangle + \frac{\gamma}{2} \sum_{n=1}^N \| \B{\hat{d}}_n \|^2 \label{opt:async_local_aggregator} \\
& \text{s.t.}
&& - \Gamma \leq r[t] \leq \Gamma, \forall t,  \nonumber\\
&&& r[t] = \sum_{n=1}^{N} \hat{d}_n[t] - \sum_{n=1}^{N} \hat{d}_n[t-1], \forall t, \nonumber\\
& \text{variables:}
&& \Gamma, \B{r}, \{\B{\hat{d}}_n\}_{\forall n}. \nonumber
\end{align}

Each prosumer performs its own local computation in asynchronous fashion including 
\begin{align} \label{eq:asyn_local_g}
	\B{w}_{g,n}^{[k+1]} = \B{z}_n^{[k]} + \gamma \B{\hat{d}}_n^{[k+1]},
\end{align}

\begin{align}
&\min && -  \langle 2 \B{w}_{g,n}^{[k+1]} - \B{z}_n^{[k]} ,\B{d}_n \rangle + \frac{\gamma}{2} \| \B{d}_n \|^2  \label{opt:asyn_local_prosumer} \\
& \text{s.t.}
&& \{ \B{d}_n, \B{e}_n, \B{x}_n, \B{y}_n \} \in \mathcal{F}_n, \nonumber 
\end{align}

\begin{align} \label{eq:asyn_local_f}
\B{w}_{f,n}^{[k+1]} = 2 \B{w}_{g,n}^{[k+1]} - \B{z}_n^{[k]} - \gamma \B{d}_n^{[k+1]}.
\end{align}

Then prosumer $n$ updates its dual variable $\B{z}_n$ and sends to the aggregator to start new iteration
\begin{align} \label{eq:asyn_z_cal}
	\B{z}_n^{[k+1]} = \B{z}_n^{[k]} + \eta_k ( \B{w}_{f,n}^{[k+1]} - \B{w}_{g,n}^{[k+1]}  ).
\end{align}

\begin{figure}[t]
	\centering
	\includegraphics[width=0.5\textwidth]{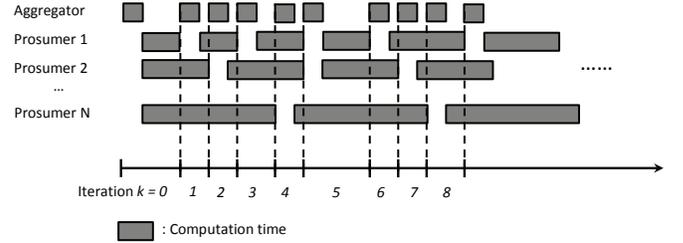}
	\caption{
		Illustration of asynchronous parallel computing. New iteration starts when any prosumer finishes its computation.
	}
	\label{fig:async_time}\vspace{-5mm}
\end{figure}

The whole procedure for asynchronous distributed algorithm can be described in Algorithm \ref{alg:asyc}. The asynchronous computation framework of Algorithm \ref{alg:asyc} is illustrated in Fig. \ref{fig:async_time}.

\begin{algorithm}[t]
	\caption{Asynchronous ADMM }\label{alg:asyc}
	\begin{algorithmic}[1]			
		\State initialize: $k = 0$, $\B{z}_n = 0, \B{\hat{d}}_n = 0, \forall n$ 		
		\Repeat 
		\State \textbf{At the aggregator:}
		\Repeat
		\State wait
		\Until{receive any update $\B{z}_n$ from prosumer $n$}
		\State \quad 1) update $\B{z}_n$ into global memory
		\State \quad 2) solve \eqref{opt:async_local_aggregator} for optimal solution $\Gamma, \{\B{\hat{d}}_{n}\}_{\forall n}, \B{r}$
		\State \quad 3) send $\B{\hat{d}}_{n}$ to prosumer $n$
		\State \quad  update global counter: $k \leftarrow k+ 1$
		\State -----------------------------------------
		\State \textbf{At any prosumer $n$}:
		\State \quad 1) calculate \eqref{eq:asyn_local_g} 
		\State \quad 2) solve \eqref{opt:asyn_local_prosumer}
		\State \quad 3) calculate \eqref{eq:asyn_local_f}
		\State \quad 4) calculate \eqref{eq:asyn_z_cal} and send to the aggregator
		\State ------------------------------------------

		\Until{a stopping criterion is met}
	\end{algorithmic}
\end{algorithm}

\section{Simulation Results} \label{sec:simulation}
In this section, we provide the numerical simulations to demonstrate the performance of the proposed model and algorithms. We test a system with $N = 100$ prosumers, and the period of scheduling is divided to $T = 24$ time slots. Each prosumer has a total daily energy consumption generated randomly around $30$ kWh, in which $30\%$ of the total demand is elastic load and can be scheduled over different time slots. The remaining demand is inelastic load, where higher energy consumption occurs from 8:00 to 22:00. The renewable energy at each prosumer is generated with a highly available amount during period from 10:00 to 20:00 \cite{denholm_overgeneration}. The energy storage system has the capacity $B_n^{max}= 4$ kWh and the initial energy level is $0.25 B_n^{max}$. The charging and discharging efficiency, $\beta_n^c = \beta_n^d = 0.9$ for all prosumers. All tests are conducted on a personal computer, in which all optimization problems in the proposed algorithms are solved using CVX \cite{cvx}.

\begin{figure}[t]
	\centering
	\includegraphics[width=0.45\textwidth]{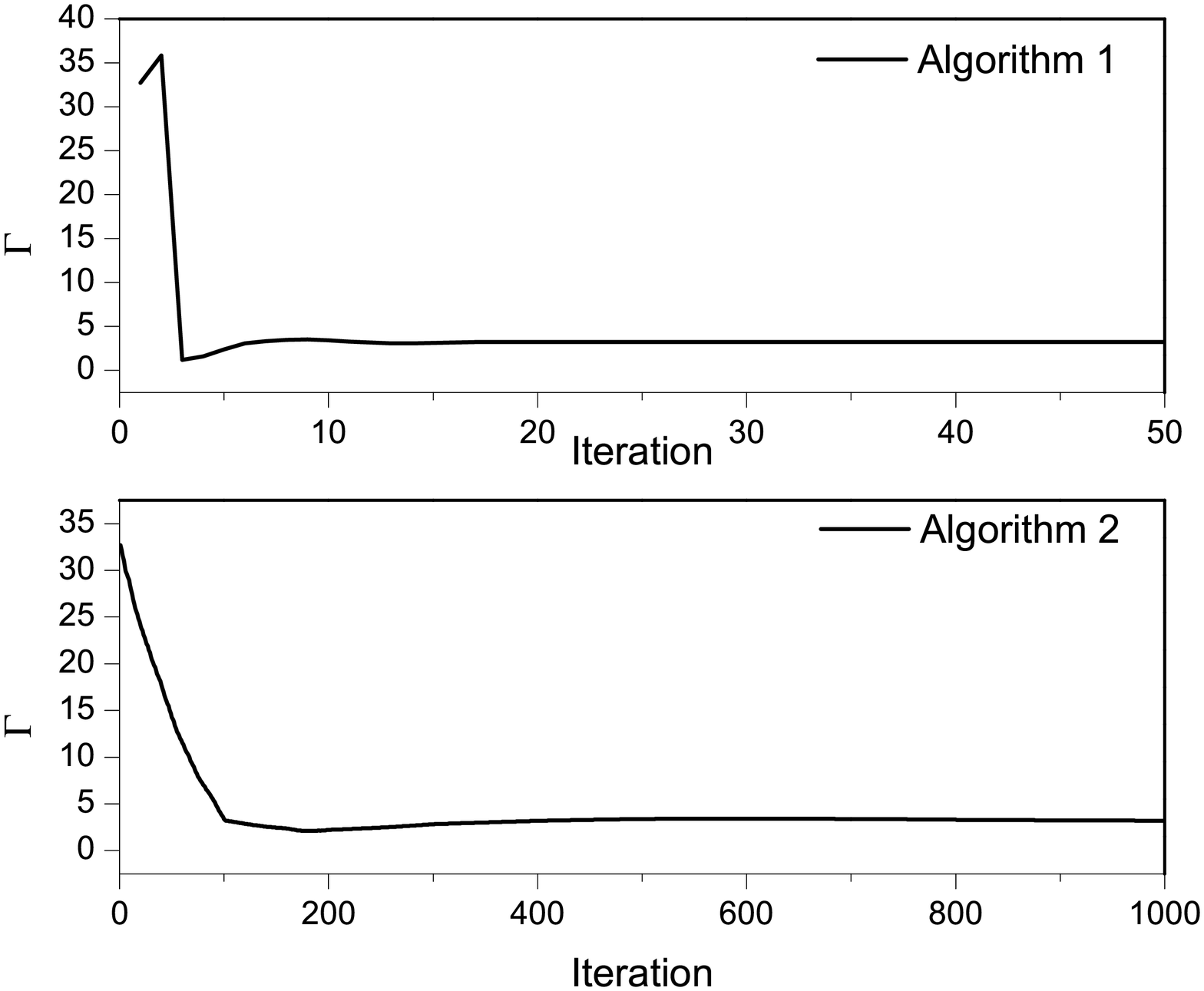}\vspace{-5mm}
	\caption{
		The convergence performance of the proposed algorithms.
	}
	\label{fig:convergence}\vspace{-5mm}
\end{figure}

To demonstrate the performance of our proposed algorithms, we show the number of iterations required for convergence in Fig. \ref{fig:convergence}. We plot the trends of the resulting objective function value over iterations. We can see that Algorithm \ref{alg:parallel} converges after about 20 iterations while Algorithm \ref{alg:asyc} needs about 400 iterations to achieve the optimal value. The faster convergence behavior of Algorithm \ref{alg:parallel} is due to the synchronous update fashion among all prosumers. However, the calculation time for each iteration in Algorithm \ref{alg:asyc} is much faster than Algorithm \ref{alg:parallel}. Particularly, the average computation time for each iteration in Algorithm \ref{alg:asyc} is 9.3 seconds, while it is 58 seconds for Algorithm \ref{alg:parallel}.

\begin{figure}[t]
	\centering
	\includegraphics[width=0.45\textwidth]{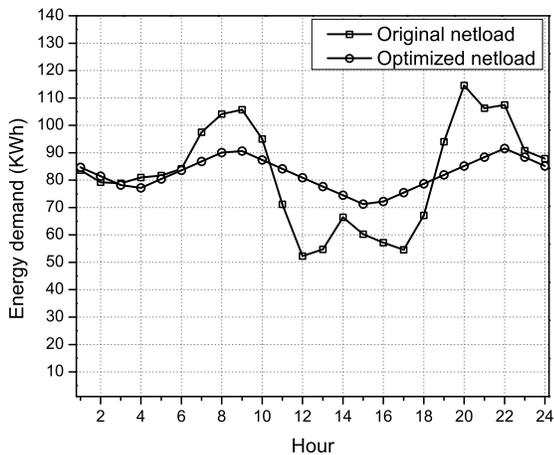}\vspace{-5mm}
	\caption{
		The net load of the system. 
	}
	\label{fig:netload}\vspace{-5mm}
\end{figure}

To illustrate the capability of our model in reducing the peak ramp, we plot the net load of the system in Fig. \ref{fig:netload} with and without the deployment of our proposed optimal scheduling. Due to the large amount of distributed renewable generation during the period from 8:00 to 20:00, prosumers use the available renewable energy to serve their load. Therefore, the total net load of the system is significantly reduced during that periods, which leads to the increased need for ramping when solar power drops off in the late afternoon (from 17:00 to 20:00) in the case of without deploying optimal scheduling algorithm. Instead, by using our proposed algorithms, the net load can be flatten over the scheduling period, and the resultant peak ramp reduces $88\%$ compared to the original net load. Note that in our model, each prosumer consumes the same amount of energy demand in two cases, but it schedules its energy consumption more efficiently to reduce the peak ramp for the overall system.

%

\section{Conclusions} \label{sec:conclusions}
In this paper, an energy consumption scheduling problem for prosumers has been studied. We first formulate a centralized optimization problem to reduce the peak ramp of the system. The global optimal scheduling can be obtained by solving the centralized problem. However, due to the privacy concerns and the distributed nature of the power system, the centralized design is difficult to implement in practice. Therefore, we propose distributed algorithms to achieve the optimal solution for the system, in which each prosumer individually schedules its energy consumption and storage profile. The first algorithm requires the synchronous update fashion from all prosumers, i.e., the new iteration starts only after all prosumers finish their calculations. On the other hand, the second algorithm can be implemented in an asynchronous fashion, i.e., the aggregator starts new iteration when any prosumer in the system finishes its calculation. The simulation results demonstrate the convergence performance of our proposed algorithms as well as the capability of our model in reducing the peak ramp of the system.

\bibliographystyle{IEEEtran}
\bibliography{IEEEabrv,SGRID}

\end{document}